\documentclass[12pt]{article}

\usepackage{amsmath,amssymb}
\usepackage{amsthm}
\usepackage{graphics,epsfig}

\theoremstyle{definition}\newtheorem{defin}{Definition}[section]
\theoremstyle{plain}\newtheorem{theo}[defin]{Theorem} 
\theoremstyle{plain} 
\theoremstyle{plain}\newtheorem{lem}[defin]{Lemma}

\title{Gravitational lensing in spherically symmetric static
spacetimes with centrifugal force reversal}
\author{
Wolfgang Hasse\thanks{
TU Berlin, Sekr.\ PN 7-1, 10623 Berlin, Germany, and Wilhelm Foerster
Observatory, Munsterdamm 90, 12169 Berlin, Germany. 
Email: astrometrie@gmx.de} 
\quad and \quad 
Volker Perlick\thanks{ Albert Einstein Institute, 14476 Golm, Germany.
(Permanent address: TU Berlin, Sekr. PN 7-1, 10623 Berlin, Germany.
Email: vper0433@itp.physik.tu-berlin.de)}}
\date{}

\begin{document}

\maketitle

%%%%%%%%%%%%%%%%%%%%%%%%%%%%%%%%%%%%%%%%%%%%%%%%%%%%%%%%%%%%%%%%%%%%%%%%

\begin{abstract}
	In Schwarzschild spacetime the value $r=3m$ of the radius
	coordinate is characterized by three different properties:
	(a) there is a ``light sphere'', (b) there is ``centrifugal
	force reversal'', (c) it is the upper limiting radius for
	a non-transparent Schwarschild source to act as a gravitational 
	lens that produces infinitely many images. In this paper we 
	prove a theorem to the effect that these three properties are 
	intimately related in {\em any\/} spherically symmetric static
	spacetime. We illustrate the general results with some examples 
	including black-hole spacetimes and Morris-Thorne wormholes.
\end{abstract}

%%%%%%%%%%%%%%%%%%%%%%%%%%%%%%%%%%%%%%%%%%%%%%%%%%%%%%%%%%%%%%%%%%%%%%%%%

\section{Introduction}\label{sec:intro}
	In a Schwarzschild spacetime with mass $m$, the horizon at the value
	$r = 2m$ of the radius coordinate plays a distinguished role. However,
	also the value $r = 3m$ is of particular interest. As a matter of fact,
	this value is characterized by three quite different properties.

	First, a geodesic with circular orbit of radius $r$ around the 
	center must be timelike for $r > 3m$, lightlike for $r = 3m$, and 
	spacelike for $r <3m$. A light ray emitted tangentially to a circle 
	of radius $r$ will go to infinity for $r > 3m$, it will stay on this
	circle for $r = 3m$, and it will go towards the center for $r < 3m$. 
	For this reason, one often refers to the surface $r = 3m$ as to a 
	``light sphere'' or a ``photon sphere''. For a detailed study of the 
	geodesics in Schwarzschild spacetime the reader may consult Chandrasekhar
	\cite{Ch}; the notion of a photon sphere is discussed from a more
	general perspective in a recent paper by Claudel, Virbhadra and
	Ellis \cite{CVE}.

	Second, an observer moving on a (non-geodesic) circular orbit of radius 
	$r$ feels a centrifugal force that is pointing in the direction of 
	increasing $r$, as in Newtonian physics, for $r > 3m$. However, for 
	$r < 3m$ the centrifugal force is pointing in the direction of decreasing 
	$r$. An observer at $r = 3m$ feels no centrifugal force at all. This
	phenomenon of centrifugal force reversal was discussed in a long series
	of articles by Marek Abramowicz with various coauthors, see e.g.
	Abramowicz and Prasanna \cite{AP}; there is also a forthcoming book
	by Abramowicz and Sonego \cite{AS}. 
	
	Third, we may fix a constant $r_1 > 2m$ and ask about the number of 
	future-pointing lightlike geodesics in the spacetime region $r > r_1$ 
	that start on a given $t$-line $\gamma$ and end at a given spacetime 
	point $p$. If we exclude the exceptional case that $p$ and $\gamma$ lie 
	on a common spatial axis through the center, this number is finite if  
	$r_1 > 3m$. However, this number is infinite if $r_1 \le 3m$. 
	Viewing $\gamma$ as the worldline of a light source and $p$ as an 
	observation event, this result admits the following interpretation. 
	A non-transparent Schwarzschild source of radius $r_1 > 3m$ 
	acts as a gravitational lens that produces finitely many images 
	whereas the number of images is infinite in the case $r_1 \le 3m$. 
	A discussion of lensing in the Schwarzschild spacetime, including  
	results on image positions on the observer's sky and on relative 
	magnitudes of the images, can be found in Virbhadra and Ellis \cite{VE}, 
	also see Frittelli, Kling and Newman \cite{FKN}.  
		
	It is an interesting question to ask whether these three properties 
	coincidentally come together in the Schwarzschild spacetime. It is the 
	purpose of this paper to demonstrate that this is not the case. More
	precisely, we are going to prove a theorem to the effect that, in any 
	spherically symmetric and static spacetime, the presence of a 
	light sphere, the occurence of centrifugal force reversal, and
	multiple imaging with infinitely many images are closely related
	phenomena.

	To work this out, we consider (3+1)-dimensional spacetimes with 
	metrics of the form
\begin{equation}\label{eq:g4}
	g = - A ( \rho )^2 dt^2 + B ( \rho )^2
	d \rho ^2 +  C ( \rho )^2 
	\big( d \vartheta ^2 + {\mathrm{sin}}^2 \vartheta d \varphi ^2 \big) 
\end{equation}
	with some strictly positive $C^2$ functions $A$, $B$ and $C$.
	Here we assume that $\vartheta$ and $\varphi$ have their usual
	range as standard coordinates on the 2-sphere $S^2$ whereas $t$ is
	assumed to range over ${\mathbb{R}}$ and $\rho$ is assumed
	to range over some open interval $\: ] \rho _{\mathrm{min}} , 
	\rho _{\mathrm{max}} [$ with $- \infty \le \rho _{\mathrm{min}} 
	<\rho _{\mathrm{max}} \le \infty$. So the topology of the
	(3+1)-dimensional spacetime manifold is ${\mathbb{R}} ^2 \times S^2$.
	
	This is the general form of a spherically symmetric and static 
	spacetime. The assumptions on $A$, $B$, and $C$ to be strictly 
	positive make sure that the Killing vector field $\partial _t$
	is timelike and that the metric has Lorentzian signature. The
	meaning of the functions $A$, $B$ and $C$ is the
	following. $A( \rho ) dt$ is the proper-time differential
	along the $t$-lines, $B( \rho ) d \rho$ is the 
	proper-length differential along the $\rho$-lines and 
	$4 \pi C( \rho )^2$ is the area of the sphere at
	constant $\rho$ and constant $t$. If the derivative 
	$C'( \rho )$ is different from zero for all $\rho \in \;
	] \rho _{\mathrm{min}} , \rho _{\mathrm{max}} [$, we may 
	transform  to a ``Schwarzschild-like radius coordinate'' $r$
	via $\rho \longmapsto r = C ( \rho )$. However, we 
	also want to include examples, such as wormholes, where 
	the derivative of $C$ does have zeros. Therefore, we stick with 
	the more general radius coordinate $\rho$. 

	It is our goal to investigate, for a point $p$ and a $t$-line
	$\gamma$ in such a spacetime, the light rays (i.e., future-pointing
	lightlike geodesics) that start on $\gamma$ and terminate at $p$.
	For any choice of $p$ and $\gamma$ we can achieve by a spatial 
	rotation that $p$ and $\gamma$ are in the hyperplane 
	$\vartheta = \pi /2$. If, in the new coordinate system, the 
	$\varphi$-coordinates of $p$ and $\gamma$ do not differ by 
	a multiple of $\pi$, then all light rays from $\gamma$ to $p$ 
	are confined to this hyperplane. In the exceptional case 
	that the $\varphi$-coordinates of $p$ and $\gamma$ do differ by a 
	multiple of $\pi$, i.e., that $p$ and $\gamma$ lie on a common radial 
	axis $P$, every light ray from  $\gamma$ to $p$ in the hyperplane 
	$\vartheta = \pi /2$ gives rise to a one-real-parameter
	family of light rays from $\gamma$ to $p$, resulting by applying
	rotations around the axis $P$. (Such a one-parameter-family 
	of light rays indicates that the observer at $p$ is seeing an 
	{\em Einstein ring\/} of the light source with worldline $\gamma$.) 
	In any case, knowledge of the light rays in the  hyperplane 
	$\vartheta = \pi /2$ will be sufficient to know all light rays. 
	For that reason we may restrict our consideration to (2+1)-dimensional 
	spacetimes with metrics of the form
\begin{equation}\label{eq:g}
	g = - A( \rho )^2 dt^2 + B( \rho )^2
	d \rho ^2 +  C( \rho )^2 d \varphi ^2 
\end{equation}
	with $t$ ranging over ${\mathbb{R}}$,
	$\varphi$ ranging over ${\mathbb{R}}$ mod $2 \pi$, and $\rho$ ranging
	over $] \rho _{\mathrm{min}} , \rho _{\mathrm{max}} [$ with 
	$- \infty \le \rho _{\mathrm{min}} <\rho _{\mathrm{max}} \le \infty$.
	
%%%%%%%%%%%%%%%%%%%%%%%%%%%%%%%%%%%%%%%%%%%%%%%%%%%%%%%%%%%%%%%%%%%%%%%%%%%%%

\section{Lightlike geodesics}\label{sec:geodesics}
	Solving the geodesic equation for the metric (\ref{eq:g}), which 
	can be done explicitly up to quadratures, is a standard exercise.
	In this section we summarize, for later convenience, the relevant 
	equations for lightlike geodesics. To that end we first observe that 
	a lightlike geodesic $\beta$ of the metric (\ref{eq:g}) admits 
	the following three constants of motion,
\begin{gather}
\label{eq:M}
	g( {\dot{\beta}},{\dot{\beta}} ) = -A(\rho)^2 \, {\dot{t}}{}^2
	+ B(\rho)^2 \, {\dot{\rho}}{}^2
	+  C(\rho)^2 \, {\dot{\varphi}}{}^2	= \, 0 \, ,
\\
\label{eq:E}
        -g( {\dot{\beta}}, \partial _t) =
	A( \rho )^2 \, {\dot{t}}
	= E \, ,
\\
\label{eq:L}
	g( {\dot{\beta}}, \partial _{\varphi} ) =
	 C(\rho)^2 \, {\dot{\varphi}}
	= L \, ,
\end{gather}
 	where an overdot denotes differentiation with respect to the 
	curve parameter. We restrict to the case $E =1$, thereby singling
	out for each geodesic a unique parametrization that is 
	future-pointing with respect to the time coordinate $t$. 

	Inserting (\ref{eq:E}) with $E = 1$ and ({\ref{eq:L}) 
	into (\ref{eq:M}) results in
\begin{equation}\label{eq:drho}
	A( \rho )^2 \, B( \rho )^2 \, {\dot{\rho}}{}^2 +
	L^2 \, V( \rho ) \, = \, 1 \, ,
\end{equation}
	where we have introduced the potential
\begin{equation}\label{eq:pot}
	V( \rho ) =  A( \rho )^2 \, C( \rho )^{-2} 
\end{equation}
	which will play the central role throughout our analysis.
	Please note that this potential is unaffected by a conformal
	change of the metric. We now divide (\ref{eq:drho}) by 
	$V(\rho )$ and differentiate the resulting equation with 
	respect to the curve parameter. After dividing by $\dot{\rho}$ 
	we find
\begin{equation}\label{eq:ddrho}
	{\ddot{\rho}} + {\dot{\rho}} {}^2 
	\big( C'( \rho ) C(\rho )^{-1} + B'( \rho ) B(\rho )^{-1} \big) 
	= - \tfrac{1}{2} \, 
	C( \rho )^2 \, B( \rho )^{-2} \, A( \rho )^{-4} \,
	V'( \rho ) \, .
\end{equation}
	Although we divided by ${\dot{\rho}}$, this equation has to
	hold, by continuity, also at points where ${\dot{\rho}} = 0$.

	Clearly, the constant map $s \longmapsto \rho (s) = \rho _0$ is a 
	solution of the differential equation (\ref{eq:ddrho}) if and only 
	if $V' ( \rho _0 ) = 0$. This is the necessary and sufficient condition 
	for a lightlike geodesic with circular orbit to exist at radius 
	$\rho _0$. From (\ref{eq:drho}) we read that, for such a geodesic,
	the constant of motion $L$ has to satisfy the equation $L^2 \, V( \rho _0 )
	=1$. A stability analysis of (\ref{eq:ddrho}) shows that a circular
	light orbit at $\rho _0$ is stable with respect to perturbations of 
	the initial condition if $V'' ( \rho _0 ) > 0$ and unstable if 
	$V'' ( \rho _0 ) < 0$.

	Also from (\ref{eq:drho}) and (\ref{eq:ddrho}) we read that along a 
	lightlike geodesic with constant of motion $L$ the radius coordinate
	has a strict local maximum (or a strict local minimum, respectively)
	at points where $L^2 \, V( \rho ) = 1 $ and $V'(\rho ) > 0$ (or 
	$V' ( \rho ) < 0$, respectively). Other extrema cannot occur along 
	lightlike geodesics with non-circular orbits.

	For later purpose we observe that (\ref{eq:L}) and (\ref{eq:drho})
	imply
\begin{equation}\label{eq:orbit}
	{\dot{\varphi}} = \frac{ L \,
	B( \rho) \, C( \rho )^{-1} }{\sqrt{
	V( \rho )^{-1}-L^2} \, } \, |{\dot{\rho}}|.
\end{equation} 
	Here we have made use of the fact that, by (\ref{eq:L}),
	${\dot{\varphi}}$ always has the same sign as $L$. 
	By integration, (\ref{eq:orbit}) yields the orbits of the light 
	rays in the $(\rho , \varphi )$-plane. 

%%%%%%%%%%%%%%%%%%%%%%%%%%%%%%%%%%%%%%%%%%%%%%%%%%%%%%%%%%%%%%%%%%%%%%%%%%%%%%%

\section{Centrifugal force reversal}\label{sec:centrifu}
	In this section we want to discuss the ``centrifugal force'' felt by
	an observer in circular motion in the metric (\ref{eq:g}). This is
	the only case that is of interest to us in this paper. For possible 
	generalizations to non-circular motions in arbitrary stationary 
	spacetimes we refer the reader, e.g., to Abramowicz, Carter and 
	Lasota \cite{ACL} and to Bini, Carini and Jantzen \cite{BCJ}. 

	On a (2+1)-dimensional spacetime with metric (\ref{eq:g}), we introduce
	the timelike vector field
\begin{equation}\label{eq:V}
	U = \frac{1}{\sqrt{1 - v^2}\, } \Big(
	A( \rho )^{-1} \, \partial _t \, \pm \,
	v \, C( \rho )^{-1} \, \partial _{\varphi} \Big)
\end{equation}
	with some constant $v \in [0,1[\,$. The integral curves of this vector
	field can be interpreted as worldlines of observers that move on
	circular orbits with constant 3-velocity $v$ (in units of the 
	velocity of light) with respect to the static observers whose
	worldines are the $t$-lines. By (\ref{eq:V}), the vector field
	U is normalized according to $g(U,U) = -1$, so its integral curves are
	parametrized by proper time. 

	In general, the integral curves of $U$ are no geodesics, i.e., 
	the 4-acceleration $\nabla _U U$ does not vanish. With respect 
	to an observer moving along an integral curve of $U$, the relative
	acceleration of a freely falling observer with a momentarily tangential 
	worldline is given by $- \nabla _U U$. In correspondence with standard
	non-relativistic terminology, this quantity could be viewed as ``inertial 
	acceleration''. We are interested in its radial component which is readily 
	calculated with the help of the identity $- g(\partial _{\rho} , 
	\nabla _U U ) = \frac{1}{2} \big( L_{\partial _{\rho}} g \big) 
	\big(\, U \, , \, U \, \big)$, where $L_{\partial _{\rho}} g$ 
	denotes the Lie derivative of $g$ with respect to the vector field 
	$\partial _{\rho} \,$. This results in 
\begin{equation}\label{eq:inert}
	- g(\partial _{\rho} , \nabla _U U ) = - A' ( \rho ) \, A(\rho )^{-1}	- 
	\frac{v^2}{2(1 - v^2)} \,  
	C( \rho )^2 \, A( \rho )^{-2} \, V'( \rho ) \, ,
\end{equation}
	with $V$ defined by (\ref{eq:pot}). 

	On the right-hand side of (\ref{eq:inert}), we interpret the 
	first term as {\em gravitational acceleration\/} and the second 
	as {\em centrifugal acceleration}. (By multiplying each of 
	those accelerations with the observer's mass we get the 
	respective ``force''.) These names are justified 
	since the first term is independent of $v$, whereas
	the second term is proportional to $v^2$ in lowest order. 
	Hence, for velocities small compared to the velocity of 
	light the centrifugal term has, indeed, the same 
	$v$-dependence as in Newtonian physics. 
  
	From (\ref{eq:inert}) we read that the sign of the centrifugal
	term is determined by the sign of $V'$. The centrifugal acceleration 
	is pointing in the direction of increasing $\rho$ at all values of 
	$\rho$ with $V' (\rho ) < 0$, and it is pointing in the direction 
	of decreasing $\rho$ at all values of $\rho$ where $V'(\rho ) > 0$. 
	In the following we are interested in the situation that $V'$ changes 
	sign at some radius $\rho _0$. In this situation we say that there is 
	``centrifugal force reversal'' at $\rho _0$. It is one of our goals
	to prove that then the gravitational field produces infinitely many 
	images for static light sources and observers at radii close to 
	$\rho _0$.

	By comparison with the preceding section we see that centrifugal 
	force reversal can occur at $\rho _0$ only if there is a circular 
	light orbit at $\rho _0$. Note, however, that the occurence of a 
	circular light orbit is not sufficient for centrifugal force reversal; 
	the potential $V$ might have a saddle. 

	The following observation is also of interest. In accordance with
	(\ref{eq:inert}), $U$ can be geodesic ($\nabla _U U = 0$) only at 
	those points where the centrifugal acceleration is exactly balanced
	by the gravitational acceleration. If the gravitational acceleration 
	is pointing in the direction of decreasing $\rho$, $A'( \rho ) > 0$, 
	this is impossible at any radius $\rho$ where $V' ( \rho ) > 0$. 
	In this sense, validity of the inequality $V' ( \rho ) > 0$ has the 
	effect that a freely falling object (with subluminal velocity) cannot 
	stay at radius $\rho$.

%%%%%%%%%%%%%%%%%%%%%%%%%%%%%%%%%%%%%%%%%%%%%%%%%%%%%%%%%%%%%%%%%%%%%%%%%%%%%%%
	
\section{Multiplicity results for light rays}\label{sec:multi}
	In the preceding sections we have emphasized the role of the potential
	$V$, defined by (\ref{eq:pot}). In particular, we have seen that the zeros 
	of $V'$ indicate circular light orbits and that the sign of $V'$ determines 
	the direction of the centrifugal force. In this section we shall state and 
	discuss a theorem that relates the occurence or non-occurence of extrema 
	(minima, maxima, or saddles) of the potential $V$ to multiple imaging. The 
	proof of this theorem will be given in the subsequent section.
	
\begin{theo}\label{theo:multi}
	Consider a $(2+1)$-dimensional Lorentzian manifold $(M,g)$
	with metric of the form $(\ref{eq:g})$, where the coordinate ranges 
	are $t \in {\mathbb{R}}$, $\varphi \in {\mathbb{R}}$ mod $2 \pi$, 
	and $\rho \in \; ] \rho _{\mathrm{min}} , \rho _{\mathrm{max}}[\,$ 
	with $- \infty \le \rho _{\mathrm{min}} < \rho _{\mathrm{max}}
	\le \infty$, hence $M \simeq {\mathbb{R}}^2 \times S^1$.
	Fix a point $p$ in $M$ 
	and an integral curve $\gamma$ of $\partial _t$ and denote the
	radius coordinates of $\gamma$ and $p$ by $\rho _1$ and $\rho _2$,
	respectively. Let $N(p, \gamma )$ be the number of future-pointing
	lightlike geodesics in $M$ that start on $\gamma$ and terminate at
	$p$, with two geodesics being identified if one is a reparametrization 
	of the other. Then the following is true for the potential $V$ defined
	by $(\ref{eq:pot})$. 
\\
	{\em (a)} If $V'$ has no zeros on the whole interval 
	$] \rho _{\mathrm{min}} , \rho _{\mathrm{max}}[\,$, then 
	$N( p , \gamma )$ is finite.
\\
	{\em (b)} If there is a $\rho _0 \in \; ] \rho _{\mathrm{min}} , 
	\rho _{\mathrm{max}}[\,$ such that $V( \rho ) \le V ( \rho _0 )$ for 
	all $\rho \in \;] \rho _{\mathrm{min}} , \rho _{\mathrm{max}}[\,$,
	then $N( p, \gamma )$ is infinite. 
\\
	{\em (c)} Assume that there is a $\rho _0 \in \; ] \rho _{\mathrm{min}} 
	, \rho _{\mathrm{max}}[\,$ with $V' (\rho _0) = 0$ such that 
	$V' ( \rho ) < 0$ for all $\rho \in \; ] \rho _{\mathrm{min}} , \rho _0[\,$
	and $V' ( \rho ) > 0$ for all $\rho \in \; ] \rho _0, \rho _{\mathrm{max}}[\,$. 
	Assume, in addition, that $\lim\limits_{\rho \to \rho _{\mathrm{min}}}
	V ( \rho ) = \lim\limits_{\rho \to \rho _{\mathrm{max}}} V ( \rho )$. Then 
	$N ( p , \gamma )$ is either zero or infinite. By keeping $\gamma$ fixed and
	moving $p$ an appropriate distance along the circle $\rho = \rho _2$ one can
	always achieve that $N(p, \gamma )$ is infinite. Moreover, by an arbitrarily
	small perturbation of the metric coefficients $A$, $B$, $C$ one
	can always achieve that $N(p, \gamma )$ is infinite for any choice of $p$
	and $\gamma$.  
\\
	{\em (d)} Assume there is a $\rho _0 \in \; ] \rho _{\mathrm{min}} 
	, \rho _{\mathrm{max}}[\,$ with $V' ( \rho _0 ) = 0$ such that 
	$V' ( \rho ) < 0$ for all 
	$\rho \in \; ] \rho _{\mathrm{min}} , \rho _{\mathrm{max}}[\,$
	with $\rho \neq \rho _0$. If $\rho _1 < \rho _0$ or 
	$\rho _2 < \rho _0$, then $N ( p , \gamma )$ is finite. If 
	both $\rho _1 \ge \rho _0$ and $\rho _2 \ge \rho _0$, then 
	$N( p , \gamma )$ is infinite. 
\\
	Part {\em (d)} implies an analogous result for the case that the inequality 
	$V' ( \rho ) > 0$, instead of $V' ( \rho ) < 0$, holds for all 
	$\rho \in \; ] \rho _{\mathrm{min}} , \rho _{\mathrm{max}}[\,$ 
	with $\rho \neq \rho _0$, simply by a coordinate transformation 
	$\rho \longmapsto - \rho$.
\end{theo}

	Every future-pointing lightlike geodesic that starts on $\gamma$
	and terminates at $p$ may be interpreted as giving an image of a 
	light source with worldline $\gamma$ for an observer at $p$. Hence, 
	in the case of part (b), part (c) and the second half of part (d)
	of the theorem there are infinitely many such images. 
	The proof of the theorem will demonstrate that in all these cases 
	the result can be strengthened in the following way. For every 
	integer $n _0 \in {\mathbb{N}}$ there are such geodesics with 
	winding numbers $n > n_0$ and $n < - n _0$. Here the {\em winding 
	number\/} of a curve $\beta : [s_1 , s_2] \longrightarrow M$ is 
	defined as the biggest integer $n \in {\mathbb{Z}}$ such that
\begin{equation}\label{eq:n}
	2 \, \pi \, n \, \le \int _{s_1} ^{s_2} {\dot{\varphi}} (s) ds 
\end{equation}
	where ${\dot{\varphi}} (s)$ is the usual shorthand notation for
	$\frac{d}{ds} \varphi \big( \beta (s) \big)$. Hence, 
	in the case of part (b), part (c) and the second half of part (d)
	of the theorem there are infinitely many images that correspond to 
	light rays winding in the positive $\varphi$-direction ($n \ge 0$) 
	and infinitely many images that correspond to light rays winding 
	in the negative $\varphi$-direction ($n < 0$). 
	 
\begin{figure}[h]
\centerline{\epsfig{figure=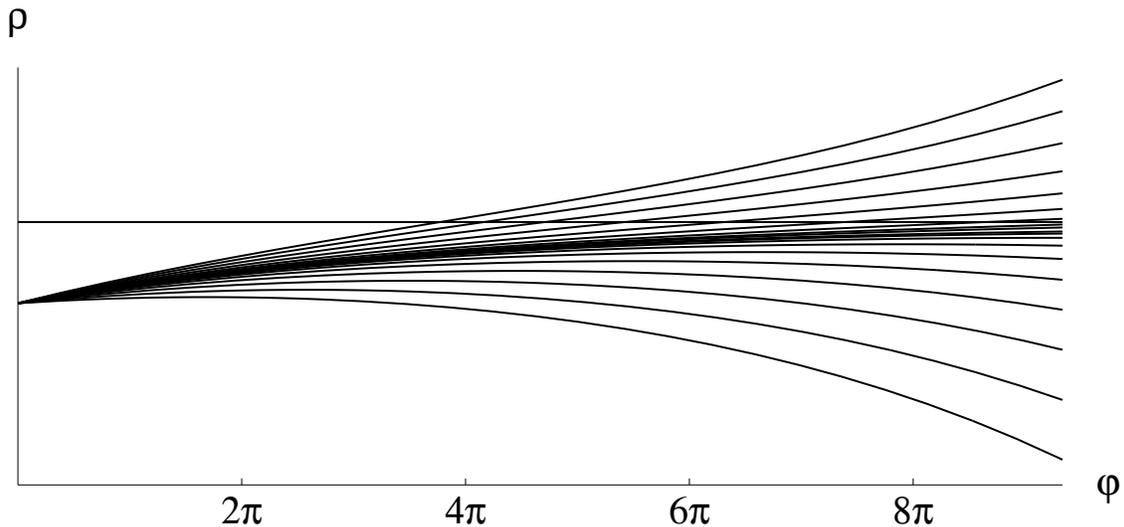, width=15cm}}
\caption{ This figure shows in a $\varphi$--$\rho$--diagram the 
	behavior of light rays near an unstable circular light orbit, or, 
	what is the same, near a local maximum of the potential $V$. We have 
	drawn light rays issuing from a point near the circular light orbit, 
	the latter being indicated by the horizontal line. For producing the 
	picture we have chosen $V(\rho ) = (a (\rho - \rho _0 ) ^2 + b)^{-1}
	$ and $B(\rho )\, C(\rho )^{-1} =c$ with some constants $a>0$, $b$ 
	and $c$. The qualitative features, however, are the same near any 
	unstable circular light orbit. As the $\varphi$--coordinate is 
	$2 \pi$--periodic, the horizontal axis should be thought of as rolled 
	into a circle. Keeping this in mind, the picture clearly illustrates 
	part (b) of Theorem~\ref{theo:multi}. } 
\label{fig:unstable}
\end{figure}

	A major value of this theorem is in the fact that any of its four 
	parts can be applied to arbitrarily small intervals 
	$] \rho _{\mathrm{min}} , \rho _{\mathrm{max}}[\,$. In
	particular, parts (b), (c) and (d) of this theorem characterize
	multiple imaging behavior near local maxima, strict local minima
	and saddles of the potential $V$. Figures~\ref{fig:unstable}, 
	\ref{fig:stable} and \ref{fig:halfstable} show the
	qualitative behavior of light rays near extrema of $V$ and may
	serve as illustrations of parts (b), (c), and (d), respectively,
	of Theorem~\ref{theo:multi}. The discussion of light rays near
	a minimum of $V$ is more subtle than near a maximum or near a
	saddle for the following reason. There is a class of spherically
	symmetric static spacetimes in which, for the constant of motion $L$ varying 
	over some interval, all light rays have periodic orbits with the
	same period, see the lower half of Figure~\ref{fig:stable}. These
	spacetimes could be viewed as lightlike analogues of the {\em 
	Bertrand spacetimes\/} discussed by Perlick \cite{Pe} which are
	characterized by periodic orbits of {\em timelike\/} geodesics.
	In such a ``lightlike Bertrand spacetime'' there are pairs of
	source and observer which cannot be connected by any light ray,
	whereas other pairs can be connected by infinitely many light rays.

\begin{figure}
\centerline{\epsfig{figure=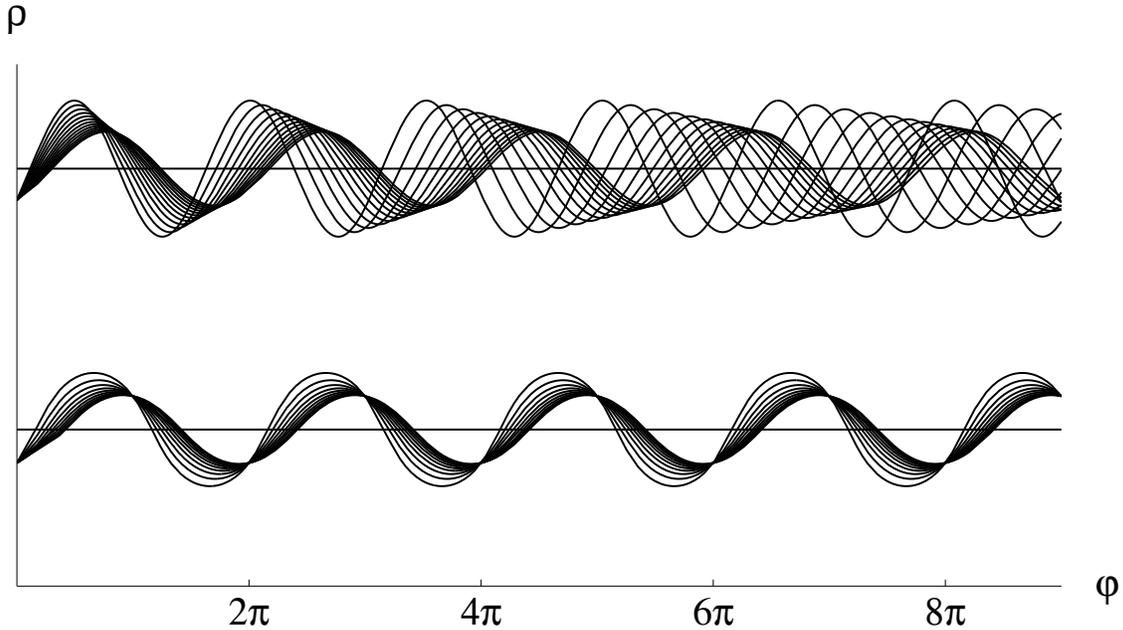, width=15cm}}
\caption{ This figure shows in a $\varphi$--$\rho$--diagram the behavior of 
	light rays near a stable circular light orbit, i.e., near a local minimum 
	of the potential $V$, for two different choices of $V$. In both cases 
	we have drawn light rays issuing from a point near the circular light orbit, 
	the latter being indicated by a horizontal line. The upper
	diagram is valid for $V( \rho )= (-a(\rho - \rho _0 )^2 + b)^{-1}$, 
	the lower diagram for $V(\rho )=a \, {\mathrm{cosh}}^2 \big( (\rho - \rho _0) 
	/b \big)$, and in both cases we have chosen $B(\rho ) \, C(\rho )^{-1} = 
	\, c \,$, with constants $a>0$, $b$ and $c$. Along any light ray close to 
	a stable circular light orbit, $\rho$ is 
	a periodic function of $\varphi$. This is true independent of the special
	form of $V$. However, the period may depend on the constant of motion 
	$L$, as in our first example, or it may be constant, as in our second 
	example where the period is equal to $2 \pi b/c$. In the first case, the 
	light rays issuing from a particular point cover each point in a neighborhood 
	of the circular light orbit infinitely often. In the second case this is 
	true only if the constant period is an irrational multiple of $2 \pi$.
	If the period is a rational multiple of $2 \pi$ (as in our picture where 
	we have chosen $b = c$), then the light rays issuing from a particular 
	point cover some points in a neighborhood of the circular light orbit 
	infinitely often whereas other points are not met at all. This is 
	obviously a highly non-generic situation. It can be destroyed by an 
	arbitrarily small perturbation of the metric coefficients in such a 
	way that the period becomes $L$-dependent (or, as an alternative, in such
	a way that the period becomes a constant but irrational multiple of $2 \pi$).
	In this sense, light rays near a {\em generic\/} stable circular light orbit 
	qualitatively behave like in the upper diagram. Some more insight can 
	be gained from studying the proof of part (c) of Theorem~\ref{theo:multi}, 
	see Section~\ref{sec:proof} below. } 
\label{fig:stable}
\end{figure}

	In a nutshell, Theorem~\ref{theo:multi} says that, in a 
	metric of the form (\ref{eq:g}), multiple imaging with infinitely
	many images occurs if and only if this metric admits a circular
	light orbit. The latter may be a local minimum, a local maximum or
	a saddle of the potential $V$. If we exclude saddles,
	then the occurence of a circular light orbit is equivalent to 
	centrifugal force reversal. As saddles are non-generic in the sense
	that they can be destroyed by an arbitrarily small perturbation of
	the metric functions, it is thus justified to summarize 
	Theorem~\ref{theo:multi} in the following way. In a generic metric
	of the form (\ref{eq:g}), centrifugal force reversal is necessary
	and sufficient for the occurence of multiple imaging with
	infinitely many images. 

\begin{figure}
\centerline{\epsfig{figure=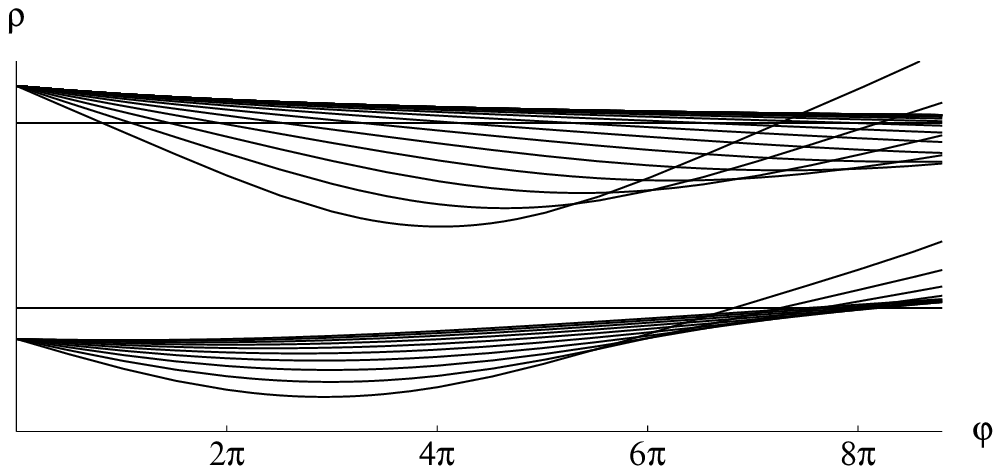, width=15cm}}
\caption{ This figure shows in a $\varphi$--$\rho$--diagram the behavior of 
	light rays near a half-stable circular light orbit, i.e., near a saddle
	of the potential $V$. For producing the picture we have chosen 
	$V(\rho )= (a (\rho - \rho _0 )^3 + b)^{-1}$ and $B(\rho )\, 
	C(\rho )^{-1}=\, c$ with constants $a>0$, $b$ and $c$. The upper 
	diagram shows light rays issuing from a point in the region 
	$\rho > \rho _0$, the lower diagram from a point in the region 
	$\rho < \rho _0$. The qualitative features are the same near any 
	saddle of $V$. }
\label{fig:halfstable}
\end{figure}

	Finally, we want to add some words of caution to prevent the reader
	from possible misinterpretations. In Theorem~\ref{theo:multi}
	we identify two lightlike geodesics if one is a reparametrization 
	of the other. This, however, does {\em not\/} imply that images are 
	identified if they are situated at the same spot on the observer's 
	sky. Running through a periodic light orbit arbitrarily often in 
	positive $\varphi$-direction (or in negative $\varphi$-direction,
	respectively) gives infinitely many images for any pair of source 
	and observer on this orbit, provided that both the source and the
	observer are ``transparent'' in the sense that they do not block
	light rays; however, all these infinitely 
	many images are situated at the same spot on the observer's sky, 
	one behind the other. So the observer will actually {\em see\/} only 
	two images, one corresponding to geodesics winding in positive 
	$\varphi$-direction and one corresponding to geodesics winding in
	negative $\varphi$-direction. This peculiar situation occurs only
	if light source and observer are on a periodic light orbit,
	so it is non-generic. In a generic situation with infinitely many
	images the images will always be situated at infinitely many different
	positions on the observer's sky.

	Also, it is worthwile to remark that, in the case of an infinite 
	sequence of images, the apparent brightness of these images necessarily 
	goes to zero. Physically, this follows from the fact that a light
	source cannot emit an infinite amount of energy. Since any detector
	has a finite sensitivity, it will register only finitely many images.
	There is a second reason why it is impossible to actually observe an 
	infinite sequence of images. As infinitely many points on the celestial 
	sphere must have an accumulation point, the limited resolution of any 
	detector implies that it is impossible to resolve all of them. 
	Therefore, a mathematical statement that there are infinitely 
	many images at infinitely many different positions on the observer's 
	sky physically only means that the observer can see arbitrarily many 
	images by choosing a detector of sufficiently high
	sensitivity and resolution (as long as the light source
	can be treated as pointlike and quantum-theoretical limits on the 
	measuring process play no role).

%%%%%%%%%%%%%%%%%%%%%%%%%%%%%%%%%%%%%%%%%%%%%%%%%%%%%%%%%%%%%%%%%%%%%%%%%%%%%%%

\section{Proof of Theorem \ref{theo:multi}}\label{sec:proof}
	We make a transformation $\, ]\rho _{\mathrm{min}} , \rho _{\mathrm{max}} [
	\; \longrightarrow \; ] u _{\mathrm{min}} , u _{\mathrm{max}} [ \; , 
	\rho \longmapsto u$ of the radius coordinate
	such that
\begin{equation}\label{eq:du}
	du = B( \rho) \, C( \rho )^{-1} \, d \rho \, ,
\end{equation}
	and we write
\begin{equation}\label{eq:f}
	f(u) = C( \rho )^2 \, A( \rho )^{-2} = V(\rho )^{-1} \, .
\end{equation}
	This puts the orbit equation (\ref{eq:orbit}) into the form
\begin{equation}\label{eq:orbitu}
	{\dot{\varphi}} = \frac{ L |\dot{u} | }{\sqrt{
	f(u) - L^2}} \, .
\end{equation} 
	For the evaluation of this equation we shall use the following
	elementary lemma.
\begin{lem}\label{lem:int}
	Let $f$ be a $C^2$ function 
	$\; ] u _{\mathrm{min}} , u _{\mathrm{max}} [ \; 
	\longrightarrow \, {\mathbb{R}}{}^+ \,$. Fix two real numbers
	$u _a$ and $u _b$ such that $u _{\mathrm{min}} < u _a < u _b <
	u _{\mathrm{max}}$ and let $K$ be a constant such that
	$K^2 \le f(u)$ for all $u \in [u_a,u_b]$.
\\
	{\em (a)} If there is a parameter value $u _0 \in [ u _a , u _b]$ with
	$f(u _0) = K^2$ and $f'(u _0 ) = 0$, then 
\begin{equation}\label{eq:infinite}
	\int _{u _a} ^{u _b} 
	\frac{ d u}{\sqrt{f(u) - K^2 \, }} 
	\; = \; \infty \; .
\end{equation}
	{\em (b)} If such a parameter value $u _0$ does not exist, then the integral
	on the left-hand side of $(\ref{eq:infinite})$ is finite.
\end{lem}
\begin{proof}
	To prove part (a) we choose a small but non-zero real number 
	$\varepsilon$ such that
\begin{equation}\label{eq:inta}
	\int _{u _a} ^{u _b} 
	\frac{d u}{\sqrt{ f(u) - K^2 \, }} \ge
	\Big| \int _{u _0} ^{u _0 + \varepsilon} 
	\frac{d u}{
	\sqrt{f(u) - K^2 \, }} \; \Big| \; .
\end{equation}
	This is possible because the integrand is positive. (If $u _0 = u _a$
	we have to choose $\varepsilon >0$; if $u _0 = u _b$ we have to 
	choose $\varepsilon < 0$. In any other case we may choose $\varepsilon$ 
	either positive or negative.) As, by assumption, $f(u_0) = K^2$ and
	$f'(u_0) = 0$, Taylor's theorem yields $f(u) = K^2 + 
	h(u) ( u - u _0 )^2$ with some continuous function $h$. By 
	the mean value theorem, there is a constant $B_0 ( \varepsilon) \in 
	{\mathbb{R}}$ such that
\begin{equation}\label{eq:Taylora}
	\Big| \int _{u _0} ^{u _0 + \varepsilon} 
	\frac{d u}{
	\sqrt{f(u) - K^2 \, }} \; \Big| = \Big|
	\frac{1}{B_0( \varepsilon )} \int _{u _0} ^{u _0 + \varepsilon} 
	\frac{d u}{u - u _0 } \; \Big| = \infty \, .
\end{equation}
	(\ref{eq:Taylora}) and (\ref{eq:inta}) demonstrate that part (a)
	of the lemma is true. --
	Under the assumptions of part(b) of the lemma, the equation 
	$f(u_0) = K^2$ can hold only for $u _0 = u _a$ with 
	$f'(u_a) > 0$ or for $u _0 = u _b$ with $f'(u_b) < 0$ (or for 
	both).  We choose a small positive $\varepsilon$ and write
\begin{equation}\label{eq:intb}
	\int _{u _a} ^{u _b} 
	\frac{ d u}{\sqrt{f(u) - K^2 \, }} = \Big(
	\int _{u_a} ^{u_a + \varepsilon} +
	\int _{u _a + \varepsilon} ^{u_b - \varepsilon} +
	\int _{u _b - \varepsilon} ^{u _b} \Big)
	\frac{d u}{\sqrt{f(u) - K^2 \, }} \, . 
\end{equation}
	If $f(u_a) \neq K^2$, the first integral on the right-hand side 
	of (\ref{eq:intb}) is certainly finite. If $f(u_a) = K^2$, Taylor's 
	theorem yields $f(u) = K^2 + C_0 ( u - u _a ) + h(u) (u - u _a )^2$ 
	with a constant $C_0 > 0$ and a continuous function $h$. Then the mean 
	value theorem guarantees the existence of a constant $A_0 (\varepsilon)
	\in {\mathbb{R}}$ such that
\begin{equation}\label{eq:Taylorb}
        \int _{u _a} ^{u _a + \varepsilon} 
	\frac{du}{
	\sqrt{f(u) - K^2 \, }} =
	\frac{A_0 (\varepsilon )}{\sqrt{C_0}} \int _{u _a} ^{u _a + \varepsilon} 
	\frac{d u}{\sqrt{ u - u_a }} \; = \; 
	\frac{2 \, A_0 (\varepsilon ) \sqrt{\varepsilon}}{\sqrt{C_0}} \, . 
\end{equation}
	So the first integral on the right-hand side of (\ref{eq:intb}) is always
	finite. By a completely analogous calculation one shows that the
	last integral on the right-hand side of (\ref{eq:intb}) is always finite.
	As the middle integral is obviously finite, this completes the proof of
	part (b) of the lemma. 
\end{proof}  
	We are now ready to prove the theorem. Having replaced $\rho$ with $u$,
	we denote the coordinates of $\gamma$ and $p$ by $( u _1 , \varphi _1 )$ 
	and $( u _2 , \varphi _2 )$, respectively. From equation (\ref{eq:orbitu})
	we read that the orbit of a lightlike geodesic remains an orbit of a 
	lightlike geodesic if it is run through in the opposite direction. 
	Therefore, it is no loss of generality if we assume that $u _1 \le u _2$. 			Moreover, we may assume $0 \le \varphi _2 - \varphi _1 < 2 \pi$.
\\[0.2cm]
	{\em Proof of part} (a) {\em of Theorem} \ref{theo:multi}: By (\ref{eq:du})
	and (\ref{eq:f}), our assumption of $V'$ having no zeros is equivalent to
	$f'$ having no zeros. We shall give the proof for the case that $V' <0$ which
	is equivalent to $f' > 0$. The case that $V'>0$ and, thus, $f'<0$ is then
	covered as well, because we are always free to change $\rho$ into $- \rho$
	and, thereby, $u$ into $-u$. If $V' <0$, we read from (\ref{eq:ddrho}) that 
	along any lightlike geodesic the coordinate $\rho$ cannot have other 
	extrema than strict local minima. By (\ref{eq:du}), the same is true for 
	the coordinate $u$. Hence, there are two classes of lightlike geodesics 
	from $(u_1, \varphi _1 )$ to $(u_2 , \varphi _2 )$: (i) those along which 
	$u$ is a strictly monotonous function, and (ii) those along which $u$ 
	has exactly one extremum, namely a local minimum. What we have to prove is that 
	both classes contain only finitely many members. For a lightlike
	geodesic of class (i), integration of the orbit equation
	(\ref{eq:orbitu}) yields
\begin{equation}\label{eq:mon}
	\varphi _2 - \varphi _1 + 2 \pi n \, = \, L \,
	\int_{u_1} ^{u_2} \frac{d u}{\sqrt{f(u) - L^2 \,}}  
\end{equation} 	  
	where $n \in {\mathbb{Z}}$ is the winding number. The possible values
	of $L$ are restricted according to $-K < L < K$, where $K = \sqrt{f(u_a(L))}$. 
	Now part (b) of Lemma~\ref{lem:int} implies that the right-hand side of 
	(\ref{eq:mon}) is bounded, i.e., (\ref{eq:mon}) can hold 
	only for finitely many integers $n \in {\mathbb{Z}}$. As the 
	right-hand side of (\ref{eq:mon}) is a strictly
	monotonous function of $L$, this proves that there are only finitely
	many values of $L$ possible for a lightlike geodesic of class (i) from 
	$( u_1 , \varphi _1 )$ to $( u _2 , \varphi _2 )$. 
	Clearly, the initial condition $u_1 , \varphi _1$ together 
	with the value of $L$ fixes a solution of (\ref{eq:drho}) and
	(\ref{eq:L}) (to be expressed in the new coordinates $(u, \varphi)$)
	uniquely up to extension. As along a geodesic of 
	class (i) $u$ cannot take the value $u _2$ more than
	once, this concludes the proof that class (i) contains
	only finitely many geodesics. -- For a lightlike geodesic of 
	class (ii), integration of the orbit equation (\ref{eq:orbitu})
	yields	
\begin{equation}\label{eq:min}
	\varphi _2 - \varphi _1 + 2 \pi n \, = \, L  
	\Big(\int_{u_a(L)} ^{u_1} + \int_{u_a(L)} ^{u_2} \Big) 
	\frac{d u}{\sqrt{f(u) - L^2 \,}}  
\end{equation} 
	where $u _a(L)$ is the minimum value of $u$ along the geodesic.
	$u _a(L)$ is related to $L$ by the equation $f(u_a(L)) = L^2$.
	Again, the possible values of $L$ are restricted according to 
	$- \sqrt{f(u_1)} < L < \sqrt{f(u_1)}$. 
	By part (b) of Lemma~\ref{lem:int}, both integrals on the 
	right-hand side of (\ref{eq:min}) remain bounded, even if $L^2$ 
	approaches its maximal value. Thus, (\ref{eq:min}) can hold only
	for finitely many integers $n$. As in the case of class (i),
	this leads to the conclusion that there are only finitely
	many values of $L$ possible for a lightlike geodesic from 
	$( u _1 , \varphi _1 )$ to $( u _2 , \varphi _2 )$. 
	As along a geodesic of class (ii) $u$ can take the value $u _2$
	at most twice, this demonstrates that class (ii) contains only 
	finitely many geodesics. 
\hfill $\square$
\\[0.2cm]
	{\em Proof of part} (b) {\em of Theorem} \ref{theo:multi}: We
	assume that the coordinate transformation $\rho \longmapsto u$ maps 
	$\rho _0$ to the value $u_0$ and we distinguish two cases: 
	(A) $u_0 \in [ u _1 , u _2 ]$, (B) $u _0 \notin [ u _1 , u _2 ]$. 
	In case (A) it is our goal to demonstrate that there is a solution 
	of (\ref{eq:orbitu}) from $(u _1 , \varphi _1 )$ to $( u_2 , \varphi _2 )$
	with winding number $n$, for any $n \in {\mathbb{Z}}$. We first
	observe that this is obviously true if $u _1 = u _2$ since,
	by assumption of case (A), this implies that $(u _1 , \varphi _1)$ 
	and $(u _2 , \varphi _2 )$ are two points on a circular light orbit
	with radius $u _0$; so we can construct the desired light rays
	simply by running through this light orbit as often as necessary. 
	Therefore, it is no loss of generality if we assume for the proof 
	in case (A) that $u _1 < u _2$. For a geodesic along which the 
	radius coordinate $u$ increases monotonically from $u _1$ to $u _2$
	integration of the orbit equation (\ref{eq:orbitu}) yields (\ref{eq:mon}).
	If we set $K = \sqrt{f(u_0)}$, the allowed values for $L$ are 
	restricted by $-K < L < K$. Evidently, the right-hand side 
	of (\ref{eq:mon}) is a strictly 
	increasing function of $L$. By Lemma~\ref{lem:int}, the integral 
	on the right-hand side of (\ref{eq:mon}) goes to infinity for 
	$|L| \rightarrow K$. As the integrand is positive, we have
	thus shown that the right-hand side of (\ref{eq:mon})
	varies monotonically from $- \infty$ to $\infty$ if $L$ 
	varies from $-K$ to $K$. Hence, for any $n \in {\mathbb{Z}}$ 
	there is an allowed value of $L$ such that 
	(\ref{eq:mon}) is satisfied. -- In case (B) we may
	assume that $u _0 < u _1 \le u _2$ because we are
	free to make a coordinate transformation $u \longmapsto
	- u $. Moreover, we may assume that $K^2 < f(u)$ for all 
	$u \in \; ] u _0 , u _1 [ \,$, 	where again $K = \sqrt{f(u_0)}$. 
	This can be achieved by replacing, if necessary, the original 
	$u _0$ with the maximal value at which the condition was violated. 
	(Please note that the theorem does not assume uniqueness of $\rho _0$
	or, what is the same, of $u_0$.) With these assumptions, we consider 
	light rays along which the radius coordinate monotonically decreases 
	from $u _1$ to some value $u _a (L) \in [ u _0 , u _1]$ and then 
	monotonically increases to $u _2$, where $f(u_a (L)) = L^2$. For 
	such a light ray integration of the orbit equation (\ref{eq:orbitu}) 
	yields (\ref{eq:min}). Here the allowed values for $L$ are restricted 
	by $K^2 \le L^2 \le (K + \varepsilon )^2$ with some positive 
	$\varepsilon$. For $|L| \rightarrow K$, which implies $u _a (L) 
	\rightarrow u _0$, both integrals in (\ref{eq:min}) go to infinity, 
	owing to Lemma~\ref{lem:int}. Thus, if $L$ varies over all allowed 
	positive (or negative, respectively) values, the left-hand side of 
	(\ref{eq:min}) varies from some positive value to
	$ + \infty$ (or from some negative value to $- \infty$, respectively).
	This implies that (\ref{eq:min}) can be satisfied for all $n \in
	{\mathbb{Z}}$ with $|n|$ bigger than some $n_0 \in {\mathbb{N}}$. 
\hfill $\square$
\\[0.2cm]
	{\em Proof of part} (c) {\em of Theorem} \ref{theo:multi}: 
	By assumption, there is a parameter value $u_0 \in \; ] u_{\mathrm{min}} ,
	u_{\mathrm{max}}[\;$ such that $f'(u) > 0$ for $u \in \; ]u_{\mathrm{min}},
	u_0[\;$ and $f'(u) < 0$ for $u \in \; ]u_0, u_{\mathrm{max}}[\;$. As 
	(\ref{eq:du}) defines $u$ only up to an additive constant, we may assume 
	that $u_0 = 0$. With $\sqrt{f(0)} = K$, the orbit equation (\ref{eq:orbitu}) 
	says that $L$ is restricted by $L^2 \le K^2$. For $L = \pm K$ we get the circular 
	light orbit at $u = u_0 = 0$; along light rays with $L^2 < K^2$, if 
	sufficiently extended, $u$ oscillates between a minimum value 
	$u _a (L)$ and a maximum value $u _b (L)$ with $f(u_a(L))=f(u_b(L)) = 
	L^2$. We are interested in light rays passing through the points 
	$(u _1 , \varphi _1 )$ and $( u _2 , \varphi _2 )$, so we must have 
	$u _a (L) \le u _1 \le u _2 \le u _b (L)$. This restricts the possible 
	values of $L$ according to  $(L_0 - \delta )^2 < L^2 \le L_0 ^2$ where 
	$L_0 \le K$ and $\delta$ are some positive constants. For any light 
	ray along which $u$ starts at $u _1$, increases monotonically to 
	$u _b (L)$, oscillates $k$ times to $u _a (L)$ and back to $u _b (L)$ 
	and finally decreases monotonically to $u _2$, integration of the orbit 
	equation (\ref{eq:orbitu}) yields
\begin{equation}\label{eq:orbitc}
	( \Delta \varphi ) ( k , L ) = \Phi (L) + k \Psi (L)
\end{equation}
	where
\begin{equation}\label{eq:Phi}
	\Phi (L) = L \Big(
	\int_{u_1 } ^{u_b (L)}  +	\int_{u_2} ^{u_b (L)} \Big) 
	\frac{du}{\sqrt{f(u) - L^2 \,}} \; .  
\end{equation} 	  
	and
\begin{equation}\label{eq:Psi}
	\Psi (L) = 2 L
	\int_{u_a (L) } ^{u_b (L)}   
	\frac{du}{\sqrt{f(u) - L^2\,}} \; .  
\end{equation} 	  
	It is our goal to prove that for any $n \in {\mathbb{Z}}$ with $|n|$ 
	sufficiently large we can choose $k$ and $L$ such that 
\begin{equation}\label{eq:delta}
	(\Delta \varphi ) (k,L) = \varphi _2 - \varphi _1 + 2 n \pi \, .
\end{equation}
	To that end, fix some $k$
	and let $L$ run over all allowed positive (or negative, respectively) 
	values. Then $( \Delta \varphi ) ( k , L )$ ranges over an interval of 
	length $\alpha + k \beta$, where $\alpha$ and $\beta$ are independent of 
	$k$ and $\beta$ is zero if and only if the function $\Psi$ is constant on 
	the interval $\;]L_0 - \delta, L_0]\,$ (or on the interval 
	$[-L_0,-L_0 + \delta[\;$, respectively). If $\beta \neq 0$, we 
	can secure the overlapping of intervals pertaining to neighboring values
	of $k$ by choosing $k$ sufficiently large, $k > \big( \Psi (L_0 )- 
	\alpha \big) / \beta$, so we can satisfy equation (\ref{eq:delta}) for any 
	$n \in {\mathbb{Z}}$ with $|n|$ sufficiently large. Now let us assume
	that $\beta =0$, i.e., that $\Psi (L)$ takes a constant value $P_0$ 
	(necessarily $P_0>0$) for all allowed values of $L$. If $2 \pi /P_0$
	is irrational, the numbers $\{ (\Delta \varphi )(k,L) | 
	k \in {\mathbb{N}}\}$ modulo $2 \pi$ are dense in the
	circle ${\mathbb{R}}$ modulo $ 2 \pi$, for any allowed value of $L$.
	Hence, if $L$ varies over an arbitrarily small interval around a positive
	(or negative, respectively) allowed value, these numbers cover the circle
	infinitely often, i.e., (\ref{eq:delta}) can be satisfied for infinitely  
	many positive (or negative, respectively) values of $n$. If $2 \pi /P_0$ is 
	rational, the numbers $\{ (\Delta \varphi )(k,L) | k \in {\mathbb{N}}\}$ 
	modulo $2 \pi$ meet only finitely many points of the circle ${\mathbb{R}}$ 
	modulo $ 2 \pi$, for any allowed value of $L$. Hence, if $L$ varies over 
	a small interval around a positive (or negative, respectively) allowed 
	value, (\ref{eq:delta}) is satisfied either for infinitely many positive
	(or negative, respectively) values of $k$ or for no such value at all,
	depending on $\varphi _2 - \varphi _1$. In the latter case one can
	obviously achieve that $N(p,\gamma )$ is infinite by moving $p$ some 
	distance along the circle $\rho = \rho _2$ (and leaving $\gamma$ fixed).
	The case $\beta =0$, i.e., the case that $\Psi$ is constant on a whole 
	interval, is indeed possible. E. g., if $f(u) = K^2 {\mathrm{cosh}} ^{-2} 
	(P u)$, we find $\Psi (L) = 2 \pi /P$ for all $L^2 < K^2$. It is also 
	possible to construct (non-analytic but arbitrarily often differentiable) 
	examples where $\Psi$ is constant on some interval but not everywhere 
	constant. The property of $\Psi$ being constant on a whole interval can
	always be destroyed by an arbitrarily small perturbation of $f$ (i.e.,
	of the metric coefficients) as can be seen from (\ref{eq:Psi}).
\hfill $\square$
\\[0.2cm]
	{\em Proof of part} (d) {\em of Theorem} \ref{theo:multi}: By assumption,
	the function $f$ has a saddle at some value $u_0$. If $u_1=u_2=u_0$, then 
	there are infinitely many light rays from $(u_1,\varphi_1)$ to 
	$(u_2,\varphi_2)$ because we can run through the circular light orbit at 
	$u_0$ as often as we like. Therefore we may exclude the case $u_1=u_2=u_0$
	for the rest of the proof. Then the assumptions imply that, along any light 
	ray from $(u_1,\varphi_1)$ to $(u_2,\varphi_2)$, $u$ cannot have other 
	extrema than strict local minima. Hence there are two classes of such light
	rays, as in the proof of part (a): (i) Those along which $u$ is monotonous
	such that (\ref{eq:mon}) holds, and (ii) those along which $u$ passes
	through exactly one local minimum at some value $u_a(L)$ such that
	(\ref{eq:min}) holds. If $u_1<u_0$ or $u_2<u_0$, then $L$ is restricted
	by $L^2 \le L_0^2$ with some $L_0^2 < f(u_0)$, so the integrals on the
	right-hand side of (\ref{eq:mon}) and (\ref{eq:min}) are bounded by
	Lemma~\ref{lem:int}. As in the proof of part (a), this implies that
	$N(p,\gamma)$ is finite. If $u_1\ge u_0$ and $u_2\ge u_0$, $L^2$ is allowed
	to vary over an interval $[K^2-\delta,K^2]$ with $K=\sqrt{f(u_0)}$.
	By Lemma~\ref{lem:int}, the right-hand side of (\ref{eq:min}) correspondingly
	varies from some positive value to $+ \infty$ for $L>0$, and it varies from
	some negative value to $-\infty$ for $L<0$. Hence, (\ref{eq:min}) can be 
	satisfied for any $n \in {\mathbb{Z}}$ with $|n|$ sufficiently large. 
\hfill $\square$	

%%%%%%%%%%%%%%%%%%%%%%%%%%%%%%%%%%%%%%%%%%%%%%%%%%%%%%%%%%%%%%%%%%%%%%%%%%%%%%%
\section{Examples}\label{sec:examples}
\subsection{Black-hole spacetimes}\label{subsec:black}
	We consider spherically symmetric and static (3+1)-dimensional
	spacetimes of the form
\begin{equation}\label{eq:gblack}
	g = - A( r )^2 dt^2 + B( r )^2 d r ^2 + 
	r^2 \big( d \vartheta ^2 + 
	{\mathrm{sin}} ^2 \vartheta \,d \varphi ^2 \big) 
\end{equation}
	where $\vartheta$ and $\varphi$ are standard coordinates on
	the 2-sphere, $t$ ranges over ${\mathbb{R}}$ and $r$ ranges 
	over an interval $\, ]r_H, \infty[\,$ with $0 < r_H < \infty$. 
	Restricting to the hyperplane $\vartheta =
	\pi /2$ gives a (2+1)-dimensional spacetime of the kind 
	considered in the preceding sections, with the Schwarzschild-like
	coordinate $r$ replacing our general radial coordinate $\rho$.

	In this situation the potential (\ref{eq:pot}) takes the form
	$V(r) = r^{-2} A(r)^2$. We assume that the Killing vector 
	field $\partial _t$ becomes lightlike, i.e., $A(r) 
	\rightarrow 0$, in the limit $r \rightarrow r_H$. This indicates 
	that there is a (Killing) horizon at $r = r_H$. In addition, we
	assume that $A(r)$ remains bounded for $r \rightarrow
	\infty$ which is true if the metric (\ref{eq:gblack}) is
	asymptotically flat. These assumptions imply that the potential
	$V(r)$ goes to zero both for $r \rightarrow r_H$ and for 
	$r \rightarrow \infty$, so the strictly 
	positive function $V$ must attain its absolute maximum somewhere 
	on the interval $\, ]r_H, \infty [ \,$. Thus, part (b) of 
	Theorem \ref{theo:multi} implies that every observer sees 
	infinitely many images of any static light source in this spacetime.
	This result should be compared with Theorem 4.3 in Claudel, Virbhadra
	and Ellis \cite{CVE} which says that, under some mathematical 
	conditions different from ours, every spherically symmetric and 
	static black hole must be surrounded by at least one ``photon 
	sphere''. In our version, there is a photon sphere at the maximum 
	of the potential $V$. Please note that $V$ may have several extrema, 
	so there may be additional photon spheres. Correspondingly, the 
	centrifugal force in such a spacetime is pointing in the direction 
	of increasing $r$ near infinity and it is pointing in the direction 
	of decreasing $r$ near the horizon; in between, it may change its 
	direction several times. 

	As a particular example we consider the Reissner--Nordstr\" om
	spacetime which is the unique spherically symmetric and static
	black-hole solution of the Einstein--Maxwell equations. The above
	result may also be illustrated with black-hole solutions of the
	Einstein--Yang-Mills--Higgs... equations. The Reissner--Nordstr\" om
	metric is of the form (\ref{eq:gblack}) with
\begin{equation}\label{eq:RN}
	A(r)^2 = B(r)^{-2} = 
	1 - \frac{2m}{r} + \frac{e^2}{r^2} \, ,
\end{equation}
	cf., e.g., Hawking and Ellis \cite{HE}, p. 156, so the potential
	$V(r)$ takes the form
\begin{equation}\label{eq:potblack}
	V(r) = \frac{1}{r^2} - \frac{2m}{r^3} + \frac{e^2}{r^4} \; . 
\end{equation}
	We restrict to the case that the constants $m$ and $e$ satisfy 
	$0 \le |e| < m$, and we let the radius coordinate $r$ range over 
	the interval $\, ] r_H, \infty[\,$ with $r_H = m + \sqrt{m^2 - e^2}$. 
	Then the Reissner--Nordstr\" om metric gives the spacetime around a
	non-rotating object with mass $m$ and charge $e$ that has undergone 
	gravitational collapse. Clearly, $V(r) \rightarrow 0$ both for \
	$r \rightarrow r_H$ and for $r \rightarrow \infty$, so the 
	above result implies that in the Reissner--Nordstr\" om 
	spacetime every observer sees infinitely many images of every static 
	light source. However, with the metric explicitly given, we can 
	strengthen this general result in the following way. From 
	(\ref{eq:potblack}) we calculate that $V$ has exactly one extremum, 
	namely a maximum at 
\begin{equation}\label{eq:r0}
	r_0 = \frac{3m}{2} + \sqrt{\frac{9m^2}{4} - 2 e^2\,} \, .
\end{equation}
	Hence, part (b) of Theorem~\ref{theo:multi} implies that
	inside any shell $r_{\mathrm{min}} < r < r_{\mathrm{max}}$ 
	that contains the radius $r_0$ every event can be reached from 
	every $t$-line by infinitely many future-pointing lightlike 
	geodesics that are completely contained in this shell. On the other
	hand, there are only finitely many such geodesics, by part (a) of 
	Theorem~\ref{theo:multi}, if the shell does not contain the radius 
	$r_0$. In the Schwarzschild case $e=0$ equation (\ref{eq:r0}) reduces
	to $r_0 = 3m$ and we find the features discussed already in the 
	introduction. -- For a more detailed discussion of light rays in 
	the Reissner-Nordstr\" om metric the reader may consult 
	Chandrasekhar~\cite{Ch}, Chapter 5, or Kristiansson, Sonego 
	and Abramowicz~\cite{KSA}.

%--------------------------------------------------------------------------	

\subsection{Wormhole spacetimes}\label{subsec:worm} 
	Morris and Thorne \cite{MT}, also see Morris, Thorne and
	Yurtsever \cite{MTY}, consider wormhole spacetimes where
	the metric has the form
\begin{equation}\label{eq:gworm}
	g = - e^{2 \Phi ( \ell )} dt^2 + d \ell ^2 + 
	r ( \ell )^2 \big( d \vartheta ^2 + 
	{\mathrm{sin}} ^2 \vartheta \,d \varphi ^2 \big) \, .
\end{equation}
	Here $\vartheta$ and $\varphi$ are standard coordinates on
	the 2-sphere, $t$ ranges over ${\mathbb{R}}$ and $\ell$ ranges 
	over all of ${\mathbb{R}}$ as well. Restricting to the hyperplane 
	$\vartheta = \pi /2$ gives a (2+1)-dimensional spacetime of 
	the kind considered in the preceding sections, with the 
	proper-length coordinate $\ell$ replacing our general radial 
	coordinate $\rho$. Morris and Thorne assume that the metric
	(\ref{eq:gworm}) is asymptotically flat for $\ell \rightarrow 
	\infty$ as well as for $\ell \rightarrow - \infty$ which means
	to require that $r ( \ell )^2 \rightarrow \infty$ whereas
	$\Phi ( \ell )$ remains bounded for $\ell \rightarrow \pm \infty$.
	As a consequence, the strictly positive potential 
\begin{equation}\label{eq:potworm}
	V(\ell ) = r(\ell )^{-2} e^{2 \Phi (\ell )}
\end{equation}
	goes to zero for $\ell \to \pm \infty$, so it must attain its 
	absolute maximum on ${\mathbb{R}}$. The (not necessarily unique) value 
	$\ell _0$ of the radius coordinate where this takes place indicates an 
	unstable circular light orbit, similar to the black-hole case. According
	to our general terminology, there is ``centrifugal force reversal'' at 
	$\ell _0$. However, we admit that in this special example our terminology 
	might be viewed as a bit misleading because neither the direction of
	increasing $\ell$ nor the direction of decreasing $\ell$ could be
	interpreted properly as ``away from the center'' everywhere. Notwithstanding 
	this semantic problem, the observation that $V$ attains its absolute 
	maximum on ${\mathbb{R}}$ makes part (b) of Theorem~\ref{theo:multi} 
	applicable. Hence, every $t$-line $\gamma$ can be joined to every point $p$ 
	by infinitely many lightlike geodesics, i.e., every Morris-Thorne wormhole 
	acts as a gravitational lens that produces infinitely many images. 
	Incidentally, this result remains true if the two asymptotically flat 
	regions are glued together (as in the lower part of Fig.1 in
	Morris and Thorne \cite{MT}); after this identification, however,
	the spacetime does not fit into our general framework because spherical
	symmetry is lost.

	More specific results are possible if we consider the special case
	that the potential (\ref{eq:potworm}) is monotonously 
	increasing on $\, ]-\infty, 0[ \,$ and monotonously decreasing on 
	$\; ]0, \infty[\,$, with a local maximum at $\ell = 0$. Then inside 
	any shell $\ell _{\mathrm{min}} < \ell < \ell _{\mathrm{max}}$ with 
	$\ell _{\mathrm{min}} < 0$ and $\ell _{\mathrm{max}} > 0$ every 
	event can be reached from every $t$-line by infinitely many 
	future-pointing lightlike geodesics that are completely contained 
	in this shell. 

%--------------------------------------------------------------------------	

\subsection{Interior Schwarzschild solution}\label{subsec:int} 
	As another illustration of our results we want to consider light
	rays in an interior Schwarz\-schild solution,  i.e., inside a spherically 
	symmetric and static material body. This is, of course, physically 
	meaningful only in the case that the body is transparent. The reader might 
	think of our interior solution as a (rough) model for a globular
	cluster.
 
	As in subsection \ref{subsec:black} we consider a spherically symmetric 
	and static spacetime of the form (\ref{eq:gblack}), but this time we
	assume that the Schwarzschild-like radius coordinate $r$ ranges over
	$\;]\, 0 \, , \, r_1 \, [\;$ with some positive constant $r_1$. We assume 
	that this spacetime metric (i) solves the Einstein field equation for a 
	perfect fluid, (ii) has a regular center, and (iii) can be continuously 
	joined to the Schwarzschild solution 
\begin{equation}\label{eq:gSchwarz}
	g = - \Big( 1 - \frac{2m}{r} \Big) \; dt^2 + 
	\Big( 1 - \frac{2m}{r} \Big) ^{-1} \, d r ^2 + 
	r^2 \big( \, d \vartheta ^2 + 
	{\mathrm{sin}} ^2 \vartheta \,d \varphi ^2 \, \big) 
\end{equation}
	at the radius $r_1$, with the pressure $p$ going to zero for $r \to r_1$. 
	Condition (ii) requires, in particular, that the metric coefficient 
	$A (r)$ remains finite for $r \to 0$, so the potential $V(r) = 
	A(r)^2 r^{-2}$ goes to infinity for $r \to 0$. Condition (iii),
	together with condition (i), requires $A'(r)$ and, thus, the 
	derivative $V'(r)$ of the potential to be continuous at $r_1$ (see, e.g.,
	Kramer, Stephani, MacCallum, Herlt \cite{KSMH}, eq. (14.2b)), i.e.,
\begin{equation}\label{eq:match}
	V'(r) \rightarrow \frac{2(3m-r_1)}{r_1^{\,4}} \qquad 
	{\mathrm{for}} \: \; r \to r_1 \, .
\end{equation}
	If $2m < r_1 < 3m$, these conditions on $V$ imply that $V'$ has to 
	change sign, i.e., that there is centrifugal force reversal, somewhere
	between $0$ and $r_1$. By Theorem~\ref{theo:multi}, this implies that 
	the gravitational field produces infinitely many images for light 
	sources and observers placed in an appropriate shell $r_{\mathrm{min}}	
	< r < r_{\mathrm{max}}$. (Either $V$ has a strict local minimum at some 
	$r_0$ such that part (c) of Theorem~\ref{theo:multi} applies to some 
	neighborhood of $r_0\,$, or $V$ is constant on some interval such that 
	part (b) of Theorem~\ref{theo:multi} applies to that interval.) 

	Already in the introduction we have discussed the known fact that a 
	Schwarzschild source of radius $r_1 \in \; ]2m,3m[\;$ produces infinitely 
	many images for any light source and any observer outside the body. The 
	analysis in this subsection demonstrates that the same is true for 
	appropriately placed light sources and observers inside the body. -- 
	For the existence of circular light orbits in an interior Schwarzschild 
	solution the reader may also consult Example~6 in Claudel, Virbhadra and 
	Ellis~\cite{CVE}. 

%-------------------------------------------------------------------------------

\section{Outlook}\label{sec:outlook}
	It is interesting to remark that some of the multiple imaging results
	presented in Theorem~\ref{theo:multi} can be proven, as an alternative,
	with the help of Morse theory. Relevant background material can be
	found in a book by Masiello~\cite{Ma}, see, in particular, 
	Theorem~6.5.6 in this book. This theorem says that, in regions of
	stationary spacetimes whose boundaries satisfy a certain ``light
	convexity'' assumption, any observer sees infinitely many images of 
	any light source. It is easy to check that, if we specialize
	to circular shells in (2+1)-dimensional spacetimes with metrics
	of the form (\ref{eq:g}), this light convexity assumption can be 
	expressed in terms of the potential (\ref{eq:pot}) in the following way.
	The shell $\rho _1 < \rho < \rho_2$ has a light convex boundary if and
	only if $V'( \rho _1 ) > 0$ and $V'( \rho _2 ) < 0$, i.e., there 
	must be centrifugal force reversal somewhere in the shell. This
	demonstrates that part (b) of Theorem~\ref{theo:multi} can be proven,
	as an alternative, with the Morse theoretical techniques detailed
	in Masiello's book.
	In this paper we were able to give elementary proofs of all results,
	using the fact that for the class of spacetimes considered the geodesic 
	equation can be explicitly integrated up to quadratures; so there was 
	no need to use ``sophisticated'' methods such as Morse theory.
	However, Morse theory could be an appropriate tool for generalizing
	our results to spacetimes which are not spherically symmetric and
	static such that an explicit analysis of the geodesic equation
	is not possible. As long as the spacetimes are stationary, the 
	above-mentioned results of Masiello~\cite{Ma} could be used as
	a basis; for Morse theory on non-stationary spacetimes
	we refer to Giannoni, Masiello and Piccione~\cite{GMP}.
	 	
%--------------------------------------------------------------------------

\end{document}